\documentclass[12pt]{iopart}
\usepackage{graphicx}
\usepackage{amsfonts}
\usepackage{amssymb}

\begin{document}
\title[Probing the last scattering surface through CMB observations]{Probing the last scattering surface through the recent and future CMB observations}
\author{Jaiseung Kim$^1$, Pavel Naselsky$^1$, Lev Krukier$^2$, Victor Datsyuk$^2$ and Oleg Datsyuk$^2$}
\address{$^1$ Niels Bohr Institute, Blegdamsvej 17, DK-2100 Copenhagen, Denmark}
\address{$^2$ Southern Federal University 105, B. Sadovaya, Rostov-on-Don, Russia}
\ead{jkim@nbi.dk}

\begin{abstract}
We have constrained the extended (delayed and accelerated) models of hydrogen recombination, by investigating associated changes of the position and the width of the last scattering surface. Using the recent CMB and SDSS data, we find that the recent data constraints favor the accelerated recombination model, though the other models (standard, delayed recombination) are not ruled out at 1-$\sigma$ confidence level.
If the accelerated recombination had actually occurred in our early Universe, baryonic clustering on small-scales is likely to be the cause of it. By comparing the ionization history of baryonic cloud models with that of the best-fit accelerated recombination model, we find that some portion of our early Universe has baryonic underdensity.
We have made the forecast on the PLANCK data constraint, which shows that we will be able to rule out the standard or delayed recombination models, if the recombination in our early Universe had proceeded with $\epsilon_\alpha\sim-0.01$ or lower, and residual foregrounds and systematic effects are negligible.
\end{abstract}


\maketitle
 
\section{Introduction}
The recombination process of cosmic plasma, which have occurred around the redshift $z_{rec}\simeq 1100$, decouples photons from baryons \cite{Modern_Cosmology,Foundations_Cosmology}. In the presence of Ly-$\alpha$ photon sources, the recombination process might proceed with delay \cite{Delayed_recombination,Constraint_Recombination,Constraint_Recombination2}, or with acceleration in the presence of baryonic clustering on small-scales \cite{Ionization_history,Baryonic_recombination,Accelerated_recombination}.
The Cosmic Microwave Background (CMB) anisotropy, which is sensitive to the ionization history of our Universe, is affected by delay or acceleration of the recombination process \cite{Delayed_recombination,Constraint_Recombination,Constraint_Recombination2,Ionization_history,Baryonic_recombination}.
The major effect on the CMB anisotropy is expected in the CMB polarization and small angular scale temperature anisotropy \cite{Constraint_Recombination,Constraint_Recombination2,Ionization_history,Accelerated_recombination}. The five year data of the Wilkinson Microwave Anisotropy Probe (WMAP) \cite{WMAP5:basic_result,WMAP5:powerspectra,WMAP5:parameter} is released and the recent ground-based CMB observations such as the ACBAR \cite{ACBAR,ACBAR2008} and QUaD \cite{QUaD1,QUaD2,QUaD:instrument} provide information complementary to the WMAP data. In near future, PLANCK surveyor \cite{PLANCK:sensitivity} is going to measure CMB temperature and polarization anisotropy with great accuracy over wide range of angular scales. In this paper, we investigate the recombination models and baryonic cloud models, using the recent CMB and SDSS data.

The outline of this paper is as follows. 
In Section \ref{recombination_model}, we investigate the extended recombination models and shows that the data constraints favor accelerated recombination models. In Section \ref{baryonic_cloud}, we discuss and constrain baryonic cloud models. 
In Section \ref{forecast}, we make the forecast on the PLANCK data constraint. In Section \ref{conclusion}, we summarize our investigation with conclusion. 

\section{Distortion on the standard ionization history}
\label{recombination_model}
The presence of extra resonance photon sources may delay the recombination of cosmic plasma \cite{Delayed_recombination,Constraint_Recombination,Constraint_Recombination2,Ionization_history}, while the presence of baryonic clustering on small scales may accelerate the recombination \cite{Ionization_history,Baryonic_recombination,Accelerated_recombination}. 
The simplest model for the production of extra resonance photons $n_{\alpha}$ is given by \cite{Delayed_recombination,Ionization_history}:
\begin{eqnarray}
\frac{d\,n_{\alpha}}{d\,t}=\epsilon_{\alpha}(z)\,H(z)\,n, \label{production_rate}
\end{eqnarray}
 where $n$ is the number density of atoms, $H(z)$ is the Hubble expansion rate at a redshift $z$, and $\epsilon_{\alpha}(z)$ is a parameter dependent on the production mechanism. Since thickness of the last scattering surface is very small in comparison to the horizon of the last scattering surface $L_{ls}$, the dependence of $\epsilon_{\alpha}(z) $ on $z$ can be parametrized as $\epsilon_{\alpha}(z_{rec}) +o(\Delta/L_{ls})$. Hence, we use the approximation $\epsilon_{\alpha}(z)\approx\epsilon_{\alpha}(z_{rec})$ through our investigation.
Though Eq. \ref{production_rate} was originally proposed for the delayed recombination, we use Eq. \ref{production_rate} to model the accelerated recombination by assigning negative values to $\epsilon_{\alpha}$ \cite{Ionization_history,Baryonic_recombination,Accelerated_recombination}.
However, it should be noted that the physical basis of the accelerated recombination (i.e. baryonic clustering on small scales) is different from that of the delayed recombination. 
We also would like to point out that the distortion on the CMB black body spectrum by extended recombination process 
($0<|\epsilon_{\alpha}|<1$) is negligible in comparison with the distortion by the re-ionization (see \cite{Antimatter} for details), and is well within the COBE FIRAS data constraint \cite{Fixen:dipole}. 

\begin{figure}[htb!]
\centering
\includegraphics[scale=.7]{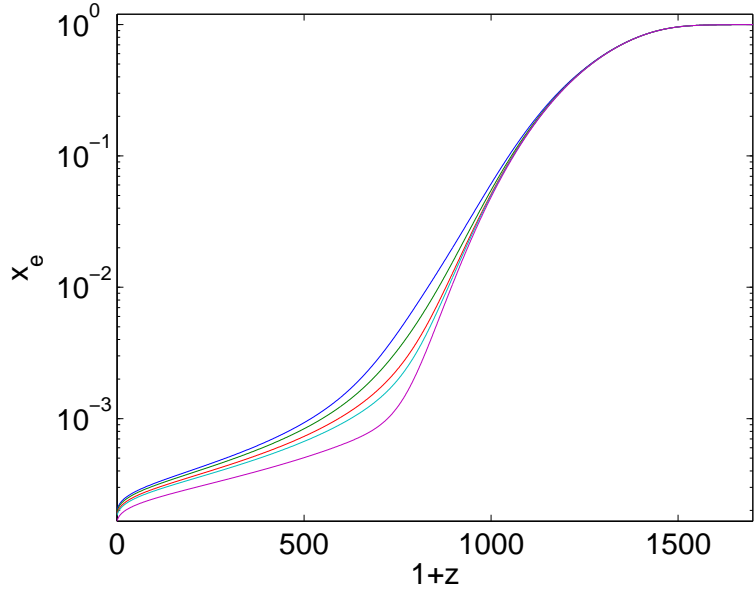}
\caption{Ionization history for $\epsilon_\alpha=0.3,\,0.1,\,0,\,-0.03,\,-0.07$ (from the highest curve to the lowest)}
\label{xe}
\end{figure}

\begin{figure}[htb!]
\centering
\includegraphics[scale=.7]{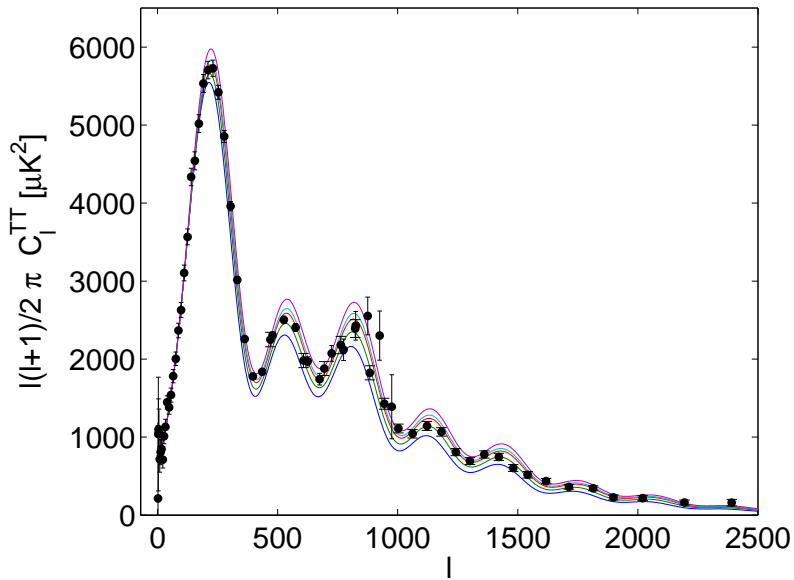}
\caption{Temperature anisotropy power spectrum for $\epsilon_\alpha=-0.07,\,-0.03,\,0,\,0.1,\,0.3$ in the descending order of Doppler peak heights. The ACBAR 2008 and the binned WMAP 5 year data are shown with error bars.}
\label{Cl_TT}
\end{figure}
By making a small modification to the \texttt{RECFAST} code \cite{RECFAST1,RECFAST2,RECFAST3}, we have computed the ionization fraction $x_e$ for various $\epsilon_\alpha$ and plotted them in Fig. \ref{xe}.
We may see that the ionization fraction $x_e$ of the recombination models $\epsilon_\alpha<0$ drops much faster than the standard model $\epsilon_\alpha=0$, while $x_e$ of the recombination models $\epsilon_\alpha>0$ drops much slower. 

\begin{figure}[htb!]
\centering
\includegraphics[scale=.7]{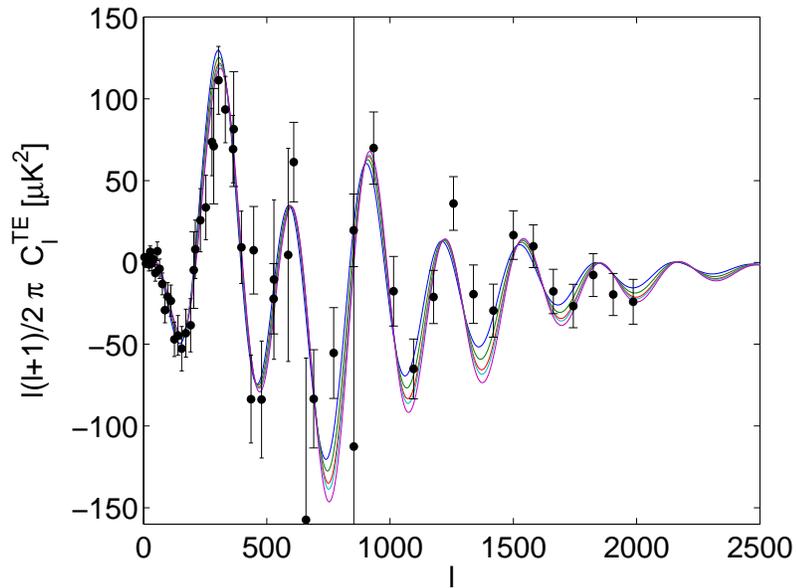}
\caption{
Temperature-E mode correlation for $\epsilon_\alpha=-0.07,\,-0.03,\,0,\,0.1,\,0.3$ in the ascending order of trough heights.
The QUaD data are shown with error bars.}
\label{Cl_TE}
\end{figure}

\begin{figure}[htb!]
\centering
\includegraphics[scale=.7]{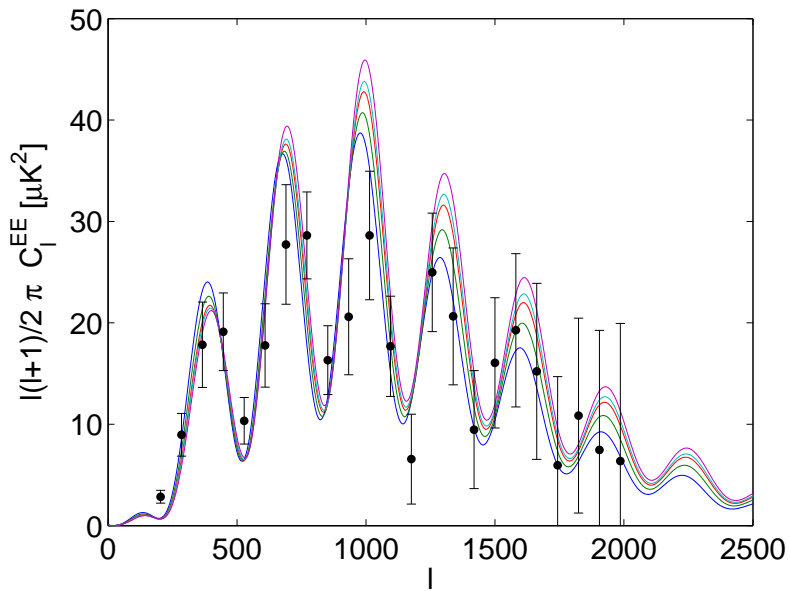}
\caption{
E mode power spectrum for $\epsilon_\alpha=-0.07,\,-0.03,\,0,\,0.1,\,0.3$ in the descending order of Doppler peak heights. The QUaD data are shown with error bars.}
\label{Cl_EE}
\end{figure}

In Fig. \ref{Cl_TT}, \ref{Cl_TE} and \ref{Cl_EE}, we show the temperature power spectra, E mode power spectra and TE correlation for various $\epsilon_{\alpha}$. It is worth noting that the location and heights of the Doppler peaks are affected by the delay and acceleration of the recombination. As shown in Fig.  \ref{Cl_TT}, the accelerated recombination models ($\epsilon_\alpha< -0.07$) are not in good agreement with the WMAP data constraint (e.g. the first Doppler peak), and the delayed recombination models ($\epsilon_\alpha> 0.3$) are ruled out \cite{Constraint_Recombination}. Hence we have investigated the recombination models ($-0.07\le \epsilon_\alpha\le 0.3$). 
For data constraints, we have used the Sloan Digital Sky Survey (SDSS) data \cite{SDSS:tech,SDSS:low_galactic,SDSS:fifth_data} and the recent CMB observations (the WMAP 5 year data \cite{WMAP5:basic_result,WMAP5:powerspectra}, the ACBAR 2008 \cite{ACBAR,ACBAR2008} and the QUaD \cite{QUaD1,QUaD2,QUaD:instrument}).
Through small modifications to the \texttt{CosmoMC} package \cite{CosmoMC}, we have included the parameter $\epsilon_{\alpha}$ of the prior distribution $-0.07\le \epsilon_{\alpha}\le 0.3$ in the cosmological parameter estimation and explored the multi-dimensional parameter space ($\Omega_{b}h^2$, $\Omega_{c}h^2$, $\tau$, $n_s$, $\log[10^{10}A_s]$, $H_0$, $\epsilon_\alpha$) by fitting the matter power spectra to the SDSS data, and the CMB anisotropy power spectra $C^{TT}_l$, $C^{TE}_l$ and $C^{EE}_l$ to the recent CMB observations (WMAP5YR + ACBAR + QUaD). 


\begin{figure}[htb!]
\centering
\includegraphics[scale=.7]{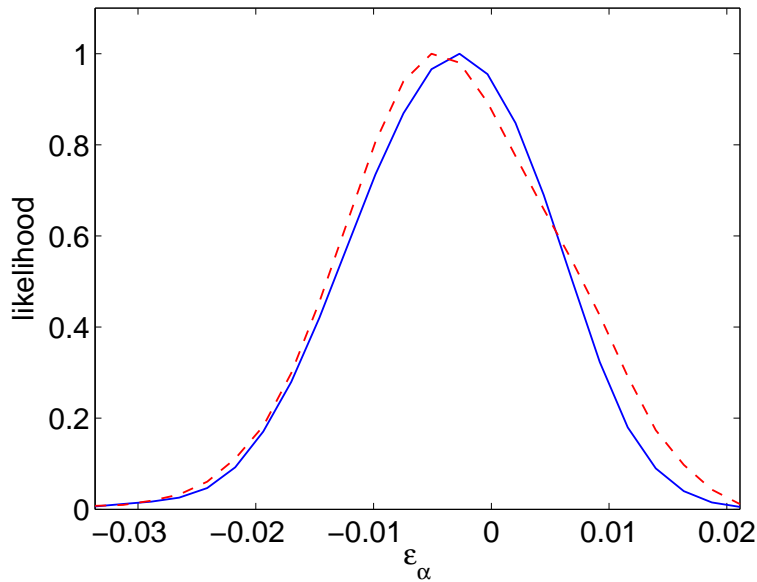}
\caption{CMB + SDSS :
the marginalized likelihood (the solid curve), the mean likelihood (the dashed curve). The normalization is chosen such that the peak value is equal to unity.}
\label{epsa}
\end{figure}
\begin{figure}[htb!]
\includegraphics[scale=.7]{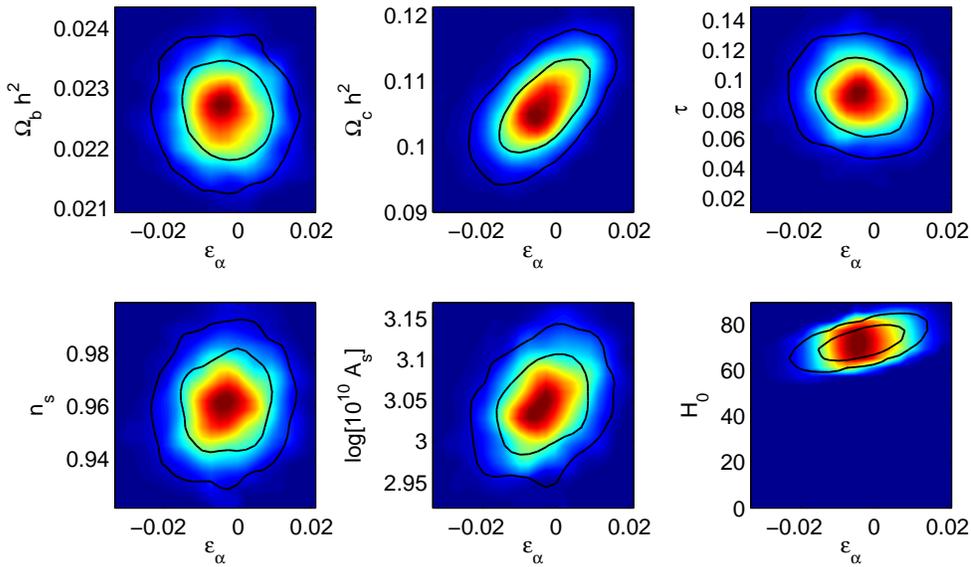}
\caption{The marginalized likelihood in the plane of $\epsilon_{\alpha}$ versus 6 basic parameters, using the CMB + SDSS constraint. Two contour lines correspond to 1$\sigma$ and 2$\sigma$ levels.}
\label{like2D}
\end{figure}
We have run the modified \texttt{CosmoMC} on a MPI cluster with 6 chains.
For the convergence criterion, we have adopted the Gelman and Rubin's ``variance of chain means'' and set the R-1 statistic to $0.03$ for stopping criterion \cite{Gelman:inference,Gelman:R1}. 
The convergence, which is measured by the R-1 statistic, is 0.0262 and 15107 chains steps are used.

In Fig. \ref{epsa}, we show the marginalized likelihood and mean likelihood of $\epsilon_{\alpha}$, given those observations.
The solid lines and dotted lines correspond to the marginalized likelihood and the mean likelihood respectively (for distinction between marginalized likelihood and mean likelihood, refer to \cite{CosmoMC}).
Though we are aware that CosmoMC does not provide a precise best-fit value, we quote the best-fit value from CosmoMC as often done in literature. 
Using the recent CMB + SDSS observation constraints, we find $\epsilon_{\alpha} = -0.00342^{+0.0185}_{-0.0237}$ at 1$\sigma$ confidence level and $\epsilon_{\alpha} = -0.00342^{+0.0214}_{-0.0305}$ at 2$\sigma$ confidence level. 
As also shown in Fig. \ref{epsa}, the current data constraints favor accelerated recombination models, though $\epsilon_{\alpha}\ge0$ is still in 1$\sigma$ confidence interval.
In Fig. \ref{like2D}, we have plotted the marginalized distribution and mean likelihoods in the plane of $\epsilon_{\alpha}$ versus \{$\Omega_{b}h^2$, $\Omega_{c}h^2$, $\tau$, $n_s$, $\log[10^{10}A_s]$, $H_0$\}. 
We summarize 1$\sigma$ constraints on cosmological parameters in Table \ref{parameter}, given CMB + SDSS data.
In comparison to the cosmological parameters estimated with the standard recombination model \cite{WMAP5:parameter}, optical depth $\tau$ is affected most, while other cosmological parameters are affected negligibly. 

\begin{table}[htb!]
\caption{cosmological parameters of $\Lambda$CDM + the extended recombination model constrained by WMAP5YR + ACBAR + QUaD + SDSS data}
\centering
\begin{tabular}{cc}
\br
parameter & 1$\sigma$ constraint\\
\mr
$\Omega_{b}h^2$  &  $0.0229^{+0.001}_{-0.0017}$\\ 
$\Omega_{c}h^2$  &  $0.1066^{+0.0114}_{-0.0121}$\\ 
$\tau$  &  $0.0930^{+0.0384}_{-0.0448}$\\ 
$n_s$ &  $0.9625^{+0.0316}_{-0.0317}$\\ 
$\log[10^{10}A_s]$ at $k_0=0.005$/Mpc &  $3.0547^{+0.0912}_{-0.1075}$\\ 
$H_0$  &  $71.7303^{+9.4648}_{-8.6242}$ km/s/Mpc\\ 
$\epsilon_\alpha$ &  $-0.0034^{+0.0185}_{-0.0237}$\\ 
\br
\end{tabular}
\label{parameter}
\end{table}

\section{Small-scale baryonic cloud models and accelerated recombination}
\label{baryonic_cloud}
As discussed in \cite{Baryonic_recombination}, the presence of the baryonic clustering in the range of very small mass scales $M\sim 10-10^5M_{\odot}$ can cause accelerated recombination. The simplest model for baryonic clustering is given by a baryonic cloud model \cite{Baryonic_recombination} as follows:
\begin{eqnarray}
\bar{\rho}_{b}=\rho_{b,in} f + \rho_{b,out} (1-f) \label{mean_rho}
\end{eqnarray}
, where $\bar{\rho}_{b}$ is the mean density of baryonic matter, and
$\rho_{b,out}$ and $\rho_{b,in}$ are the baryonic density of baryonic clouds and intercloud regions respectively, and $f$ is the total volume fraction of intercloud regions.
Denoting the baryoninc density contrast between intercloud region and clouds by $\xi=\rho_{b,in}/\rho_{b,out}$,
it may be easily shown that 
\begin{eqnarray}
\rho_{b,in} = \frac{\xi \bar{\rho}_{b}}{1+f (\xi-1)},\;\;\;\;\rho_{b,out} &=& \frac{\bar{\rho}_{b}}{1+f (\xi-1)}. \label{rho}
\end{eqnarray}
On the other hand, in the presence of baryonic clouds, there arises diffusion from clouds into intercloud regions, washing out density contrast. The characteristic scale of such diffusion is close to the Jean's length, $R_J\sim c_s \eta_r$, where $c_s$ is the baryonic speed of sound and $\eta_r=\int cdt/a(t)$ is the conformal time corresponding to the recombination time. 
Baryonic clouds of mass scales $M\sim 10-10^5M_{\odot}$ has length scale $R_{\alpha}<R_J<R<R_{\gamma}$, where 
$R_{\alpha}$ and $R_{\gamma}$ are the mean free path of resonance photons and CMB photons respectively.
Hence we may neglect baryonic diffusion.
Free electron density of clouds and intercloud regions, which we denote by $n_{e,out}$ and $n_{e,in}$ respectively,
are given by
\begin{eqnarray}
n_{e,in} = x_{e,in} \left(1-\frac{Y_{p,in}}{2}\right) n_{b,in},\;\;\;
n_{e,out} = x_{e,out} \left(1-\frac{Y_{p,out}}{2}\right) n_{b,out}, \label{ne}
\end{eqnarray}
where $x_{e,out}$ and $x_{e,in}$ are the ionization fraction of clouds and intercloud regions, and 
$Y_{p}$ and $n_{b}$ are the mass fraction of ${}^4 \mathrm{He}$ and baryon number density respectively. The ionization fraction $x_{e}$ is given by $n_{e}/(n_{e}+n_{p})$, where $n_{p}$ is the total number density of free protons and protons trapped in nucleus.
Since scales of baryonic clouds are smaller than the mean free path of CMB photons, the effective ionization fraction, which CMB anisotropy is sensitive to, is given by the mean ionization fraction:
\begin{eqnarray}
\langle x_{e}\rangle =\frac{\langle n_{e}\rangle}{\langle n_{b}\rangle}/\left( 1-\frac{\langle Y_p\rangle}{2}\right), \label{mean_xe}
\end{eqnarray}
where the mean value of free electron density is given by \cite{Baryonic_recombination}:
\[\langle n_{e} \rangle = n_{e,in} f +n_{e,out} (1-f).\]
Using Eq. \ref{mean_rho}, \ref{ne} and Eq. \ref{mean_xe}, we may show that the effective ionization fraction has the following relation to the ionization fraction of clouds and intercloud regions:
\begin{eqnarray}
\langle x_{e}\rangle = x_{e,in} G_{in} + x_{e,out} G_{out}, 
\end{eqnarray}
where 
\begin{eqnarray}
 G_{in} &=& \frac{\xi f}{1+f(\xi-1)}\left(\frac{1-Y_{p,in}/2}{1-\langle Y_p\rangle/2} \right),\\
  G_{out} &=& \frac{1-f}{1+f(\xi-1)}\left(\frac{1-Y_{p,out}/2}{1-\langle Y_p\rangle/2} \right). \label{G}
\end{eqnarray}
We would like to remind that the effective ionization fraction $\langle x_{\mathrm {e}}\rangle$ is not necessarily equal to the ionization fraction of homogeneous baryonic model (i.e. $\xi=1$), because the recombination process has non-linear dependence on baryonic density.

\begin{figure}[htb!]
\centering
\includegraphics[scale=.6]{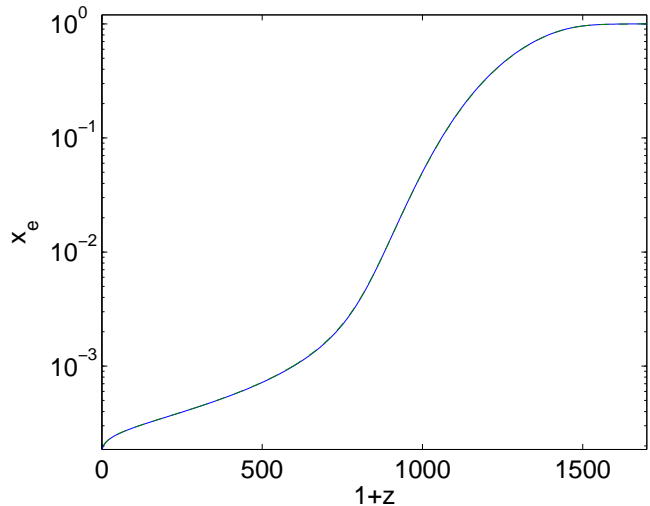}
\caption{Ionization history of the accelerated recombination model ($\epsilon_{\alpha}=-0.0034$) and the baryonic cloud model 
($f=0.019$, $\xi=0.038$): the solid curve shows the ionization history of the accelerated recombination model and
the dashed curve shows ionization history of a baryonic cloud model. Two curves are visually identical.}
\label{recfast}
\end{figure}
As presented in the previous section, the CMB data constraints with or without SDSS data favor the accelerated recombination models. Given certain values of $f$ and $\xi$, we can compute ionization history (i.e. $x_{\mathrm{e}}(z)$) of cloud and intercloud region, using the \texttt{RECFAST} with the respective values of $Y_p$ and $\Omega_\mathrm{b}$ for cloud and intercloud region.
Exploring two dimensional parameter space ($f$, $\xi$), we have fitted the ionization history of baryonic cloud models to the best-fit accelerated recombination model.
In Fig. \ref{recfast}, we show the ionization fraction $x_e$ as a function of redshift $z$. The solid curve shows the ionization fraction $x_e$ of the accelerated recombination model ($\epsilon_{\alpha}=-0.0034$), while the dashed curve shows the mean ionization fraction of the fitted baryonic cloud model ($f=0.019$, $\xi=0.038$). As shown in Fig. \ref{recfast},
two curves in Fig. \ref{recfast} are visually indistinguishable and  the fitting errors are within the accuracy of \texttt{RECFAST} \cite{Rico}.
\begin{table}[htb!]
\centering
\caption{Best-fit baryonic cloud model}
\begin{tabular}{cccccc}
\br
Constraints &$\epsilon_\alpha$ & $f$  & $\xi$&$\rho_{b,in}/\bar{\rho}_{b}$&$\rho_{b,out}/\bar{\rho}_{b}$\\ 
\mr
CMB + SDSS & -0.0034 &  0.019 & 0.038 & 0.04 & 1.02\\ 
\br
\end{tabular}
\label{best_fit}
\end{table}

Provided that $\epsilon_{\alpha}=-0.0034$, we find that baryonic clouds with baryonic density $1.02\,\bar{\rho}_{\mathrm {b}}$ occupy $\sim 98\%$ of the total volume in our early Universe, while intercloud regions with baryonic density $0.04\, \bar{\rho}_{\mathrm {b}}$ occupy $\sim 2\%$.
In Table. \ref{best_fit}, we summarize the best-fit values of $f$, $\xi$, $\rho_{b,in}/\bar{\rho}_{b}$ and $\rho_{b,out}/\bar{\rho}_{b}$.

\section{Forecast on the PLANCK data constraint}
\label{forecast}
In this section, we forecast the PLANCK data constraint on $\epsilon_{\alpha}$.
We have assumed FWHM=$10'$, $\Delta T/T=2.5$ (Stokes I) and $\Delta T/T=4$ (Stokes Q\&U) \cite{PLANCK:sensitivity,PLANCK:mission}. The pixel size is assumed to be equivalent to HEALPix \cite{HEALPix:Primer,HEALPix:framework} pixellization of Nside=1024 (Res 10). Assuming isotropic noise, we compute noise power spectrum as follows:
\begin{eqnarray}
N_{l}=N_0\,\Delta \Omega\,\e^{l^2 \sigma^2} \label{Nl}
\end{eqnarray}
, where $\sigma=\mathrm{FWHM}/\sqrt(8\ln 2)$, $N_0$ is noise variance per pixel and $\Delta \Omega$ is the solid angle of a single pixel \cite{Modern_Cosmology}. 

\begin{figure}[htb!]
\centering
\includegraphics[scale=.6]{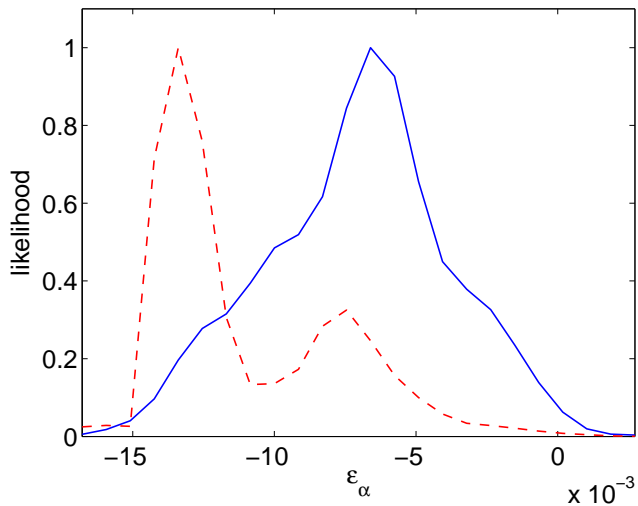}
\caption{Planck mock data ($\epsilon_\alpha=-0.01$) : the marginalized likelihood (the solid curve), the mean likelihood (the dashed curve). The normalization is chosen such that the peak value is equal to unity.}
\label{epsa_planck2}
\end{figure}

We have obtained the forecast by feeding the PLANCK mock data and the SDSS data to the \texttt{CosmoMC}. The Planck mock data are generated by drawing spherical harmonic coefficients ($2\le l\le 1500$) of signal and noise respectively from Gaussian distributions, whose variance are equal to the CMB power spectra and Eq. \ref{Nl} respectively. We have obtained the CMB power spectra by using \texttt{CAMB} \cite{CAMB} with the modified \texttt{RECFAST} and $\epsilon_\alpha=-0.01$. In the Fig. \ref{epsa_planck2}, we show the marginalized likelihood (the solid curve) and the mean likelihoods (the dashed curve) of $\epsilon_{\alpha}$, given the Planck mock data. 

Given the mock data, the \texttt{CosmoMC} finds $-0.015<\epsilon_\alpha<-3.5\times 10^{-6}$ and $-0.017<\epsilon_\alpha<0.0023$ at 1$\sigma$ and 2$\sigma$ level respectively, which is similar to the forecast by \cite{Constraint_Recombination2}.
While the \texttt{CosmoMC} forecast in Fig. \ref{epsa_planck2} is made with the mock data of $\epsilon_\alpha=-0.01$, we need to make forecast for $\epsilon_\alpha$ of various values.  
However, it is not practically feasible to run \texttt{CosmoMC} for various $\epsilon_\alpha$ because of time usually required for running \texttt{CosmoMC}.
Hence, we have estimated 1$\sigma$ interval for various $\epsilon_\alpha$ via the Fisher matrix \cite{Modern_Cosmology,Math_methods}, 
which is given by \cite{Modern_Cosmology,Math_methods}:
\begin{eqnarray}
\mathcal F_{ij}&=&\langle-\frac{\partial^2(\ln\mathcal{L})}{\partial \lambda_i\partial \lambda_j}\rangle\label{Fisher_matrix}\\
&=&\frac{1}{2} \mathrm{Tr}[\frac{\partial \mathbf S}{\partial \lambda_i} (\mathbf S+\mathbf N)^{-1}\frac{\partial \mathbf S}{\partial \lambda_j}(\mathbf S+\mathbf N)^{-1}],\label{Fisher_matrix1}
\end{eqnarray}
where $\lambda_i$ denotes the parameters to be estimated.
Evaluated at the maximum of the likelihood, the square root of diagonal element of the inverse Fisher matrix yields the marginalized $1-\sigma$ error on the parameter estimation \cite{Modern_Cosmology,Math_methods}.
The likelihood function for temperature and polarization anisotropy may be written as follows \cite{WMAP1:parameter_method}:
\[\mathcal{L}=\frac{1}{(2\pi)^{\frac{N}{2}}{\left|\mathbf S+\mathbf N\right|}^{\frac{1}{2}}}\exp[-\frac{1}{2}\Delta {(\mathbf S+\mathbf N)}^{-1}\Delta],\]
where $N$ is the number of data, $\Delta$ is a data vector, $\mathbf S$ is a signal covariance matrix and $\mathbf N$ is a noise covariance matrix. We expand anisotropy map in spherical harmonics so that our data vector $\Delta$ consists of spherical harmonic coefficients $a^{T}_{lm}$ and $a^{E}_{lm}$. In general, the computation of Eq. \ref{Fisher_matrix1} for the high resolution whole-sky observation such as the PLANCK observation is not possible on the modern computer at this moment. 
To facilitate the computation, we have made a few approximations. We assume very effective foreground cleaning with no need for sky masking and uniform instrument noise so that 
the signal and noise covariance matrices are block-diagonal and do no depend on spherical harmonic index $m$.
In such configuration, signal and noise covariance matrices are given by:
\begin{eqnarray*}
\mathbf S&=&\mathrm{diag}\left(\left(\begin{array}{cc}C^{TT}_{2}&C^{TE}_{2}\\C^{TE}_{2}&C^{EE}_{2}\end{array}\right)
,\cdots,\overbrace{\left(\begin{array}{cc}C^{TT}_{l}&C^{TE}_{l}\\C^{TE}_{l}&C^{EE}_{l}\end{array}\right)
,\cdots,\left(\begin{array}{cc}C^{TT}_{l}&C^{TE}_{l}\\C^{TE}_{l}&C^{EE}_{l}\end{array}\right)}^{2l+1}\right),
\end{eqnarray*}
and
\begin{eqnarray*}
\mathbf N=\mathrm{diag}\left(\left(\begin{array}{cc}N_{2}&0\\0&N_{2}\end{array}\right)
,\cdots,\overbrace{\left(\begin{array}{cc}N_{l}&0\\0&N_{l}\end{array}\right)
,\cdots,\left(\begin{array}{cc}N_{l}&0\\0&N_{l}\end{array}\right)}^{2l+1}\right). 
\end{eqnarray*}
Taking into account the repeating pattern of diagonal blocks, we may show 
\begin{eqnarray}
\fl\frac{1}{2} \mathrm{Tr}[\frac{\partial \mathbf S}{\partial \lambda_i} (\mathbf S+\mathbf N)^{-1}\frac{\partial \mathbf S}{\partial \lambda_j}(\mathbf S+\mathbf N)^{-1}]=\frac{1}{2} \mathrm{Tr}[\mathbf L\,\frac{\partial \tilde{\mathbf S}}{\partial \lambda_i} (\tilde{\mathbf S}+\tilde{\mathbf N)}^{-1}\frac{\partial \tilde{\mathbf S}}{\partial \lambda_j}(\tilde{\mathbf S}+\tilde{\mathbf N})^{-1}],\label{Fisher_matrix2}
\end{eqnarray}
where
\begin{eqnarray*}
\mathbf L=\mathrm{diag}\left(\left(\begin{array}{cc} 5&0\\0&5\end{array}\right),\cdots,
\left(\begin{array}{cc} 2l+1&0\\0& 2l+1\end{array}\right)
\right),
\end{eqnarray*}
\begin{eqnarray*}
\tilde{\mathbf S}=\mathrm{diag}\left(\left(\begin{array}{cc} C^{TT}_{2}& C^{TE}_{2}\\C^{TE}_{2}&
C^{EE}_{2}\end{array}\right),\cdots,\left(\begin{array}{cc} C^{TT}_{l}& C^{TE}_{l}\\C^{TE}_{l}&
C^{EE}_{l}\end{array}\right)\right),
\end{eqnarray*}
\begin{eqnarray*}
\tilde{\mathbf N}=\mathrm{diag}\left(\left(\begin{array}{cc} N_{2}&0\\0&N_{2}\end{array}\right),\cdots,
\left(\begin{array}{cc} N_{l}&0\\0&N_{l}\end{array}\right)
\right).
\end{eqnarray*}
The right hand side of Eq. \ref{Fisher_matrix2} can be easily computed in reasonable amount of time even for Post-PLANCK whole-sky observations. We chose six basic parameters ($\Omega_b h^2$, $\Omega_c h^2$, $h$, $A_s$, $n_s$) plus $\epsilon_\alpha$ for estimation parameters and set the values of basic six parameters to the WMAP best-fit values \cite{WMAP5:parameter,WMAP1:parameter_method,WMAP3:parameter}.
Due to non-linear dependence of CMB power spectra on parameter $\lambda_i$, we resorted to numerical differentiation to obtain ${\partial \tilde{\mathbf S}}/{\partial \lambda_i}$. 
The numerical derivatives are obtained by computing the following:
\begin{eqnarray}
\frac{C_l(\lambda_i+\frac{1}{2} \Delta \lambda_i)-C_l(\lambda_i-\frac{1}{2} \Delta \lambda_i)}{\Delta \lambda_i}, \label{derivative_Cl}
\end{eqnarray}
where we have set $\Delta \lambda_i/\lambda_i=10^{-3}$ for six parameters.
For the given set of parameter $\lambda_i$, we have computed $C^{TT}_{l}$, $C^{TE}_{l}$, $C^{EE}_{l}$, using \texttt{CAMB} \cite{CAMB} with the modified \texttt{RECFAST}. 
While we have made the forecast for various $\epsilon_\alpha$ in the range of $-0.07\le\epsilon_\alpha\le-0.3$, we find that Eq. \ref{derivative_Cl} has numerical instability for $\Delta \epsilon_\alpha$ of very small values.
Hence, we fixed $\Delta \epsilon_\alpha$ to be $-0.001$ instead of setting $\Delta \epsilon_\alpha/\epsilon_\alpha=10^{-3}$.

\begin{figure}[htb!]
\centering
\includegraphics[scale=.48]{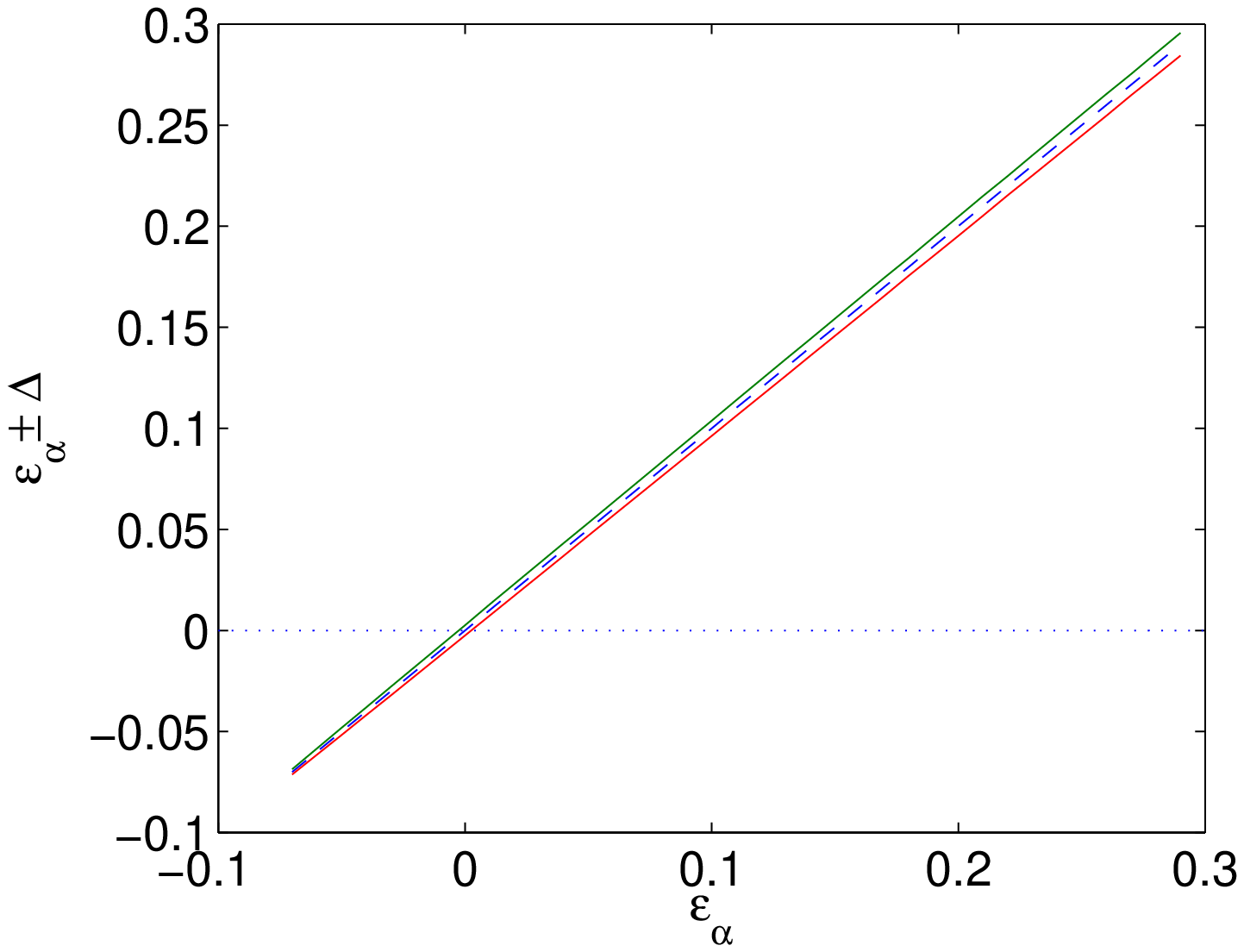}
\includegraphics[scale=.48]{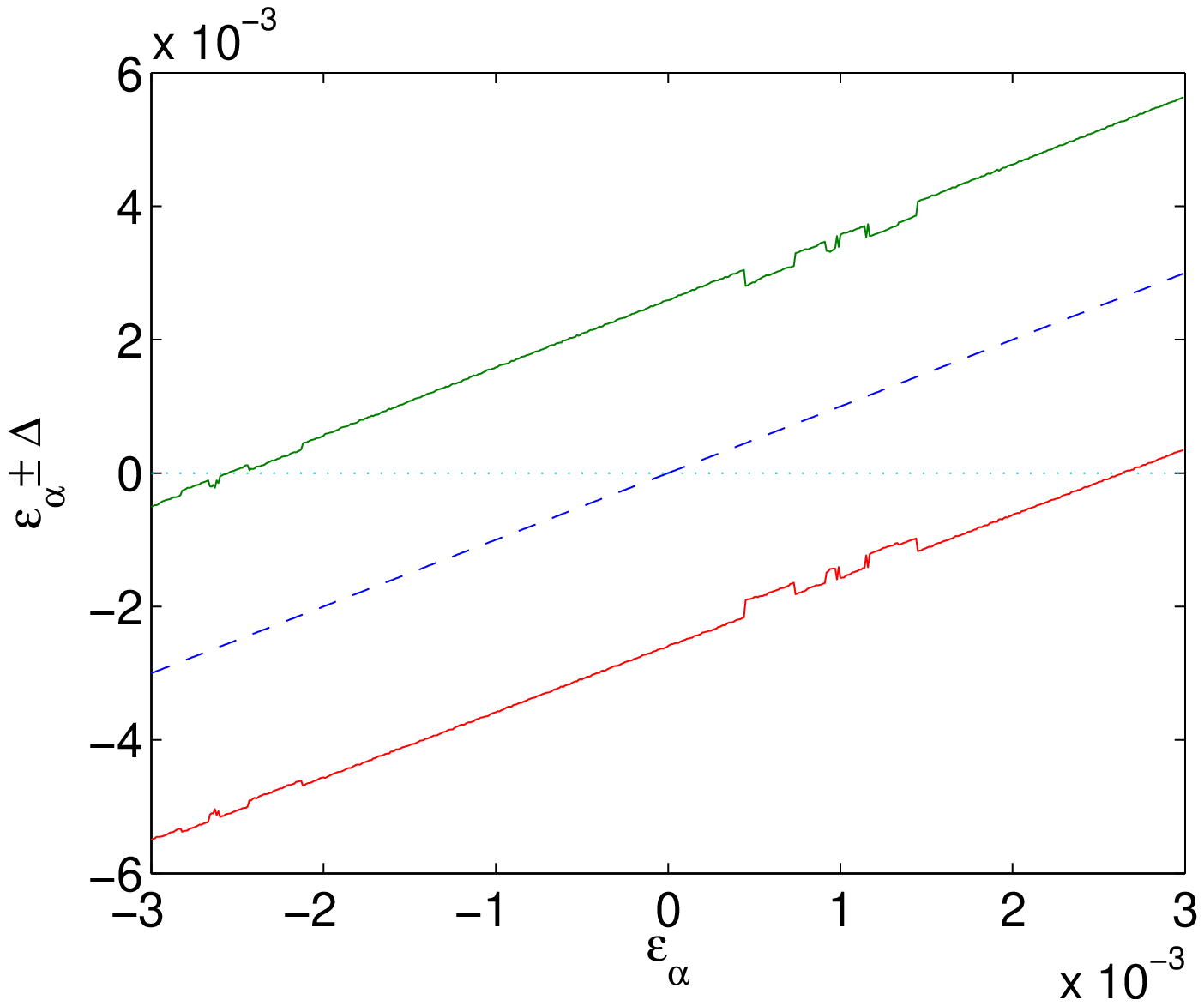}
\caption{forecast on $\epsilon_{\alpha}$ estimation from the PLANCK data}
\label{epsa_planck}
\end{figure} 
The left plot in Fig. \ref{epsa_planck} shows the estimation error for $-0.07 \le \epsilon_\alpha\le 0.3$, while the right one shows the estimation error in the vicinity of $\epsilon_\alpha=0$.
Two solid lines denote the boundary of 1$\sigma$ confidence interval and the dashed line shows the central values of 1$\sigma$ interval, which are set to the true values of $\epsilon_\alpha$.
We find that 1$\sigma$ error tends to increase with decreasing $\epsilon_\alpha$ and approaches $\sim 0.005$. We also find that the 1$\sigma$ error forecast by Fisher matrix method is smaller than the 1$\sigma$ forecast by \texttt{CosmoMC}. 
The discrepancy between \texttt{CosmoMC} forecast and Fisher matrix forecast is attributed to the deviation of parameter likelihood from Gaussian distribution. 
While Fisher matrix forecast is based on the marginalized likelihood with Gaussian approximation, \texttt{CosmoMC} forecast estimates 1$\sigma$ and 2$\sigma$ interval from mean likelihood \footnote{For Gaussian distribution, marginalized likelihood and mean likelihood are identical.}. 
However, we can learn from the Fisher matrix forecast that 1$\sigma$ error for $-0.7\le \epsilon_\alpha \le 0.3$ are at the same order of magnitude, which provides complementary information to the \texttt{CosmoMC} forecast.
For \texttt{CosmoMC} forecast as well as Fisher matrix forecast, we have neglected residual foregrounds and systematic effects including anisotropy of instrument noise and beam asymmetry, which will be additional effective sources of noise.
Hence, our forecast should be regarded as more of a lower limit on the estimation variance.

\section{Discussion}
\label{conclusion}
We have investigated the extended recombination models, using the recent CMB and SDSS data. 
We find that the data constraints favor accelerated recombination models, though other recombination models (standard, delayed recombination) are not ruled out at 1$\sigma$ confidence level.
By comparing the ionization history of baryonic cloud models with the best-fit accelerated recombination model, 
we have constrained the baryonic cloud models, from which we find that our early Universe might have slight overdensity of baryonic matter $\sim1.02\,\bar{\rho}_{\mathrm {b}}$ for $\sim 98 \%$ of total volume and underdensity of baryonic matter $\sim 0.04\,\bar{\rho}_{\mathrm {b}}$  for the rest of space. 
The origin of primordial baryonic clouds, if exists, might be associated with inhomogeneous baryogenesis  \cite{inhomogeneous_baryogenesis} in our very early Universe.

While we have constrained baryonic cloud models indirectly by fitting the ionization history, more accurate constraint will be obtained only when CMB power spectra of the baryonic cloud models are fitted directly to the data. Once we get enough evidence of accelerated recombination from the upcoming PLANCK data, we plan to constrain baryonic clouds models directly.

\ack
We are grateful to the anonymous referees for thorough reading and comments, which leads to improvements of this paper in many aspects. We acknowledge the use of the Legacy Archive for Microwave Background Data Analysis (LAMBDA), ACBAR 2008 and
QUaD 2008 data. This work made use of the \texttt{CosmoMC} package \cite{CosmoMC} and \texttt{RECFAST} \cite{RECFAST1,RECFAST2,RECFAST3}.
The numerical analysis was performed on the HPC cluster facility of the Southern Federal University of the Russia. 
This work is supported by FNU grants 272-06-0417, 272-07-0528 and 21-04-0355.

\section{References}
\bibliographystyle{unsrt}
\bibliography{/home/tac/jkim/Documents/bibliography}
\end{document}